\documentclass[11pt]{article} \usepackage{latexsym} \usepackage{graphicx} \usepackage{enumerate} \usepackage{delarray}
\usepackage{tabularx} \usepackage{bm} \usepackage{amssymb}
\usepackage[small,labelsep=period,figurename=Fig. ]{caption}
\usepackage{amsmath}
\makeatletter
\def\@biblabel#1{}
\makeatother

\begin{document}

\title{{\LARGE A model of financial contagion with variable asset returns may be replaced with a simple threshold model of cascades}} 

\author{Teruyoshi Kobayashi$^{1*}$ \\
\\
$^1$ Graduate School of Economics, Kobe University, \\
 2-1 Rokkodai, Nada, Kobe 657-8501, Japan. \\
E-mail: kobayashi@econ.kobe-u.ac.jp \\
Tel, fax: +81-78-803-6692}

%
\date{}\maketitle
\begin{abstract}
  I show the equivalence between a model of financial contagion and the threshold model of global cascades proposed by Watts (2002). The model financial network comprises banks that hold risky external assets as well as interbank assets. It is shown that a simple threshold model can replicate the size and the frequency of financial contagion without using information about individual balance sheets.
\end{abstract} \thispagestyle{empty} \newpage \pagenumbering{arabic}

\section{Introduction}
 Bank default is contagious. The failure of a single bank can spread through financial networks, generating default cascades. Over the past few years, many researchers in various fields of natural and social sciences, such as physicists, ecologists and economists, have been addressing the question of how to prevent financial contagion (e.g., Nier et al., 2007, Soramaki et al., 2007, Gai and Kapadia, 2010, Gai et al., 2011,  Lenzu and Tedeschi, 2012, Kobayashi, 2013, Kobayashi and Hasui, 2014).
 
 However, there is no wide agreement among researchers about how to construct a model of financial contagion. Different research groups use different models, which makes it difficult to establish a consensus about policy implications.\footnote{See Lorenz et al. (2009) and Upper (2011) for a survey of the literature.}
 
 In this letter, I show the equivalence between a model of financial contagion and the widely-used threshold model of global cascades proposed by Watts (2002).\footnote{See, for example, Dodds and Watts (2004), Gleeson and Cahalane (2007), Watts and Dodds (2007) and Gleeson (2013). } Basically, financial network models require researchers to construct bank balance sheets. The influence of a bank failure is then examined by sequentially undermining the interbank assets of the lenders. Second-round defaults occur if the number of defaulted borrowers among total borrowers exceeds a certain threshold. 

  This mechanism is closely related to that of the Watts model of cascades. I show that there is no need to construct bank balance sheets as long as the ``shadow'' threshold of default is appropriately defined in accordance with the volatility of assets.

\section{The models}

\subsection{A model of financial contagion} 
 The model of financial contagion used in this paper is an extended version of Gai and Kapadia (2010). The essential difference is that I take into account stochastic fluctuations in the value of external assets.

\begin{figure}
\begin{center}
\includegraphics[width=8cm,clip]{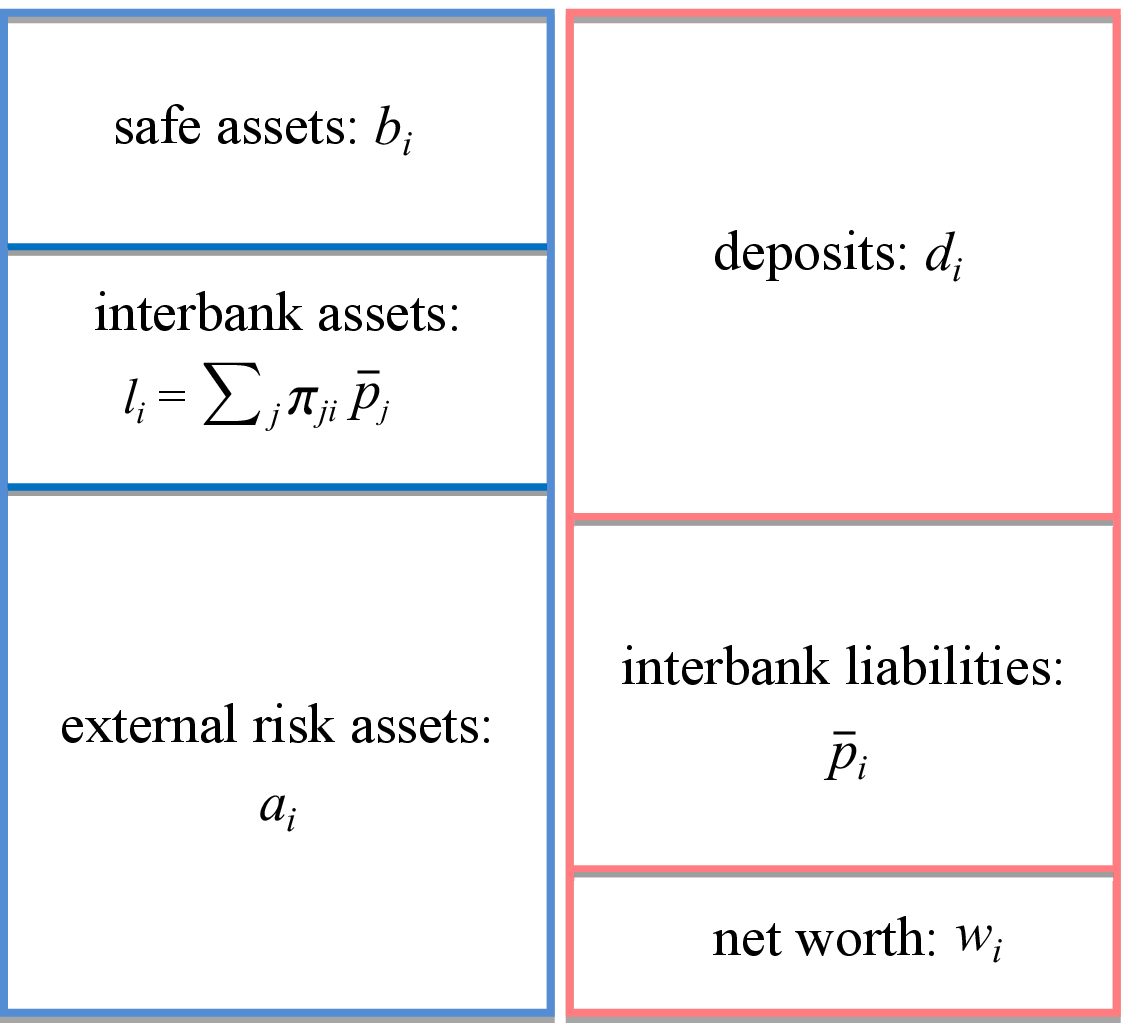}
\caption{A typical bank's balance sheet.}
\end{center}
\end{figure}

 A typical bank's balance sheet is illustrated in Figure 1. There are $N$ banks in the financial market. Bank $i$ holds risky external assets, $a_i$, interbank assets, $l_i$, and riskless assets, $b_i$. On the liability side, there are deposits, $d_i$, interbank liabilities, $\bar{p}_i$, and net worth, $w_i$. The balance-sheet condition implies that $a_i+l_i + b_i$ $= d_i + \bar{p}_i+ w_i$.

Banks are connected to each other by lending and borrowing. The existence of a lending-borrowing relationship is expressed as a \textit{link} or an \textit{edge}.
In network theory, the number of outgoing links is called \textit{out-degree}, while the number of incoming links is called \textit{in-degree}. The direction of links represents the flow of funds at the time of initial lending. 

The amount of bank $i$'s borrowings from bank $j$ is expressed as $\pi_{ij}\bar{p}_i$, where $\pi_{ij}$ denotes the relative weight of bank $i$'s borrowings from $j$, and thereby $\sum_{j\neq i} \pi_{ij}=1,\; i = 1,\ldots , N$. 
The amount of bank $i$'s total interbank assets is given by $ l_i  = \sum_{j\neq i}^N \pi_{ji}\bar{p}_j$.

 Bank $i$ will default if
 \begin{equation}
 \bar{p}_i > \sum_{j\neq i}\pi_{ji}{p}_j + \tilde{a}_i + b_i-d_i,
 \end{equation}
where $\tilde{a}_i$ and $p_j$ stand for the \textit{ex-post} values of external assets and interbank liabilities, respectively. It should be pointed out that deposits, $d_i$, are reserved because they are senior to interbank assets. 

Provided that there is no loss in interbank assets, a bank will default with probability $\delta$ due to a loss of external assets. $\delta$ indicates the probability of \textit{fundamental defaults}, which is assumed to be common across banks. 

If bank $j$ fails, then bank $j$\rq{}s creditors lose all of the credits they extended to bank $j$. Some of these creditors may fail due to the loss of their interbank assets. Accordingly, the creditors of the creditors of bank $j$ may fail as well, because they in turn lose the credits they extended to the failed banks. 

Given the size of each interbank asset, the total sizes of interbank assets and liabilities are determined by the structure of the interbank network. To ensure that the probability of fundamental defaults is the same across banks, the relative size of external assets to net worth is fixed. 
If bank $i$ has so many incoming links that its liability side is bigger than its tentative total assets, $l_i+a_i+\bar{b}_i$,  where $\bar{b}_i$ denotes the tentatively assigned riskless assets, then riskless assets are added to adjust the asset side. Otherwise, deposits are imposed to adjust the liability side.\footnote{Notice that this adjustment does not affect the probability of fundamental default.}  
After such adjustments, the capital ratio, $w_i/(l_i+a_i+b_i)$, may differ across banks while the tentative capital ratio, $w_i/(l_i +a_i + \bar{b}_i) \equiv \gamma$, is common.
The tentative ratio of total interbank assets to total assets, $ l_i/(l_i+a_i+\bar{b}_i) \equiv \theta_{l}^i $, is allowed to vary across banks.

\subsection{The Watts model of global cascades}

 Let us summarize the Watts' (2002) threshold model of cascades. In this model, each node in the network takes one of two states: ``flipped'' or ``not flipped''. The network is undirected.

Let $k_i$ and $m_i$ denote node $i$'s degree and the number of its flipped neighbors, respectively.  The algorithm of cascading behavior in the Watts model is that 
\[
  \begin{cases}
    \mbox{node $i$ flips if}\:\: m_i > \phi_i k_i,  \\
    \mbox{node $i$ does not flip otherwise},
  \end{cases}
  \]
  where $\phi_i\in[0,1]$ is the threshold of flipping for node $i$. This means that the threshold number of flipped neighbors above which node $i$ will flip is $\lfloor \phi_i k_i \rfloor$, where $\lfloor x \rfloor$ denotes the floor function that returns the maximum integer smaller than $x$. 

 \section{Equivalence between a financial network model and a threshold model}

 Here, I show the equivalence between the financial network model and a modified version of the threshold model. Recall that the \textit{ex-post} value of external assets, $\tilde{a}_i$, is determined stochastically. Let  $\Delta a_i \equiv \tilde{a}_i-a_i$ be the return of external assets.  
  
   Let us suppose that the amount of individual interbank lending is common. Then, if $\Delta a_i= 0$, bank $i$ will fail if the fraction of its defaulted borrowers exceeds $\lfloor w_i/l_i \rfloor$. Note that there is no possibility of contagious default if $w_i > l_i$, as long as there is no loss in external assets. 
 
 If $\Delta a_i \neq 0$, on the other hand, the threshold for the fraction of defaulted borrowers depends on the realization of asset returns. If a bank earns positive (negative) returns from its external assets, it becomes more resilient against (susceptible to) contagious default.
 
Let $\tilde{\phi}_i$ be the ``shadow'' threshold of default for bank $i $. It follows that
\begin{eqnarray}
\tilde{\phi}_i &=&  \frac{w_i + \Delta a_i}{l_i} \nonumber \\
          &=& \frac{\gamma}{\theta_{l}^i} +  \frac{ \Delta a_i}{l_i}, \;\; \mathrm{for}\;\; i \in \{i\:|\: l_i>0\}.
\end{eqnarray}
Here, $\theta_l^i$ is treated as a parameter. 
Thus, if asset returns, $\Delta a_i$, follow a distribution of mean zero and variance $\sigma_i^2$, then the shadow threshold $\tilde{\phi}_i$ follows a distribution of mean $ \frac{\gamma}{\theta_{l}^i}$ and variance  $\left( \frac{\sigma_i}{l_i}\right)^2$. More generally, the $p.d.f.$ of $\tilde{\phi}_i$, defined as $f_i(\cdot)$, is  given by
\begin{equation}
 f_i(x) = l_i\cdot g_i(x\cdot l_i- \gamma l_i/\theta_{l}^i), \;\; i \in \{i\:|\: l_i>0\}.
\end{equation}
where $g_i(\cdot)$ is the p.d.f. of $\Delta a_i$.

 Let us assume that asset returns follow a normal distribution with mean zero. Let $\tilde{z} \equiv F^{-1}(\delta )$, where $F^{-1}$ is the inverse $c.d.f.$ of the standard normal distribution. The standard deviation of asset returns, $\sigma_i$, such that the probability of fundamental default becomes $\delta$ is given as
 \begin{eqnarray}
  \sigma_i &=& \frac{-w_i}{\tilde{z}}, \;\;  \forall \;i,  \nonumber \\
          &=& \frac{-\gamma l_i}{\theta_{l}^i \tilde{z}},  \;\; \mathrm{for}\;\; i \in \{i\:|\: l_i >0\}.
 \end{eqnarray} 
 It follows from Eqs. (2) and (4) that the shadow threshold $\tilde{\phi}_i$ follows a normal distribution with mean $\gamma/\theta_{l}^i$ and standard deviation $-\gamma/(\theta_{l}^i\tilde{z})$.\footnote{Recall that $\tilde{z}$ takes a negative value.}
 
  In this way, the balance sheet-based model of financial contagion shown above can be expressed as a simple threshold model. Intuitively, the shadow threshold will become smaller as the returns of external assets take a lower value, meaning that the bank becomes more susceptible to default contagion. Those banks that have a negative value of $\tilde{\phi}_i$ will fail at the beginning, which corresponds to the case of fundamental defaults in the model of financial contagion.\footnote{Note that $\mathrm{Prob}(\tilde{\phi}_i <0)$ = $\mathrm{Prob}(\frac{\gamma}{\theta_{l}^i}-\frac{\gamma y}{\theta_{l}^i\tilde{z}} <0 )$ = $F(y < \tilde{z})$ = $\delta$, where $y$ is a random variable from the standard normal distribution.}

\begin{figure}
\begin{center}
\includegraphics[width=15cm,clip]{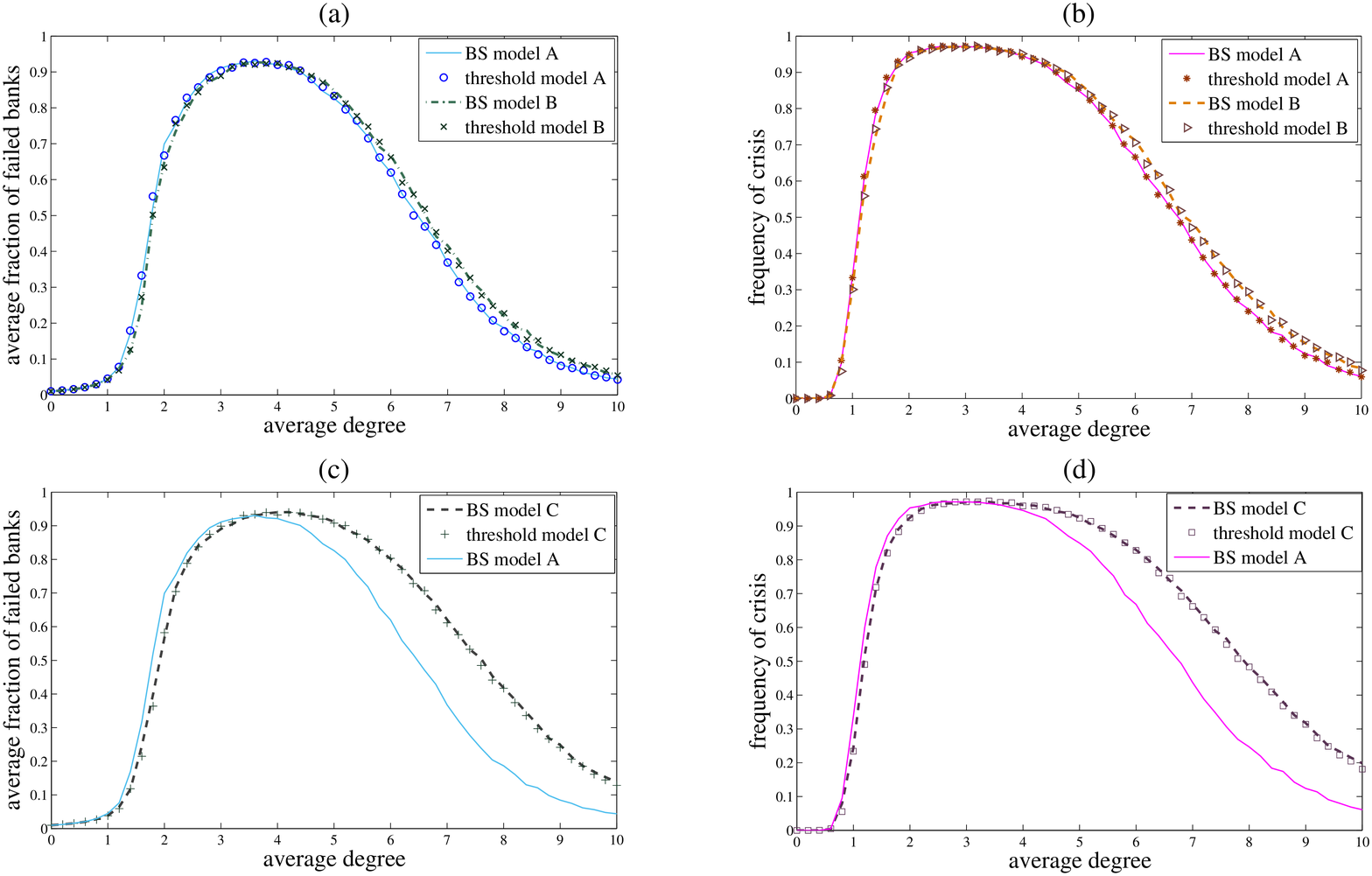}
\caption{Comparison between balance sheet-based models (BS models) and the corresponding threshold models. \textit{Note:} ``Crisis'' is defined as a situation in which at least 5\% of banks go bankrupt. BS model A: the baseline model. BS model B: the ratio of interbank assets to total assets is distributed on [.2, .4]. BS model C: the size of individual interbank assets is distributed on [.2, 1.8].}
\end{center}
\end{figure}

 I consider three variants of the model: Cases A, B and C. In Case A, or the baseline model, the size of individual interbank assets is fixed at unity, and  $\theta_{l}^i = \theta_l = .3$ for $i \in \{i\:|\: l_i >0\}.$ In Case B, $\theta_l^i$ is assumed to be uniformly distributed on [.2, .4]. In Case 3, the size of individual interbank asset is uniformly distributed on [.2, 1.8].
 
 In Figure 2, a balance sheet-based contagion model, called ``BS model'', is compared with the corresponding threshold model. With respect to BS model C, the corresponding threshold model is modified as follows: 
 \[
  \begin{cases}
    \mbox{node $i$ flips if}\:\:  \mu_i > \phi_i,  \\
    \mbox{node $i$ does not flip otherwise},
  \end{cases}
  \]
 where $\mu_i \in \: [0,1]$ is the sum of the flipped neighbors' weights. It should be pointed out that $\mu_i$ is not necessarily equal to $m_i/k_i$ unless identical weights are given to all the neighbors.  

  Given the average degree, Erd$\ddot{\mathrm{o}}$s-R\'{e}nyi (directed) random networks are created 20 times, and asset returns and shadow thresholds are independently generated 1000 times for each network structure. Other parameters are as follows: $N = 1000$, $\gamma = .1$  and $\delta = .01$. 
 
  Figure 2 reveals that in all cases, an appropriately defined threshold model can replicate the size and frequency of financial contagion created by the BS model.

\section{Conclusion and discussion}
The methods shown in this letter will enable us to analyze financial contagion without detailed information about every single balance sheet.  What is needed is the distribution of balance-sheet components among banks.  Fluctuations in asst returns and cross-sectional differences in various balance-sheet parameters can be incorporated into the threshold model.

\section*{Acknowledgments}
 I thank Kohei Hasui for his excellent research assistance.  Financial support from KAKENHI 25780203 and 24243044 is gratefully acknowledged.


\begin{thebibliography}{99}

\bibitem{}
Dodds, P.S., Watts, D. J.,
2004.
Universal behavior in a generalized model of contagion. 
\newblock{Phys. Rev. Lett.} 92, 218701.

\bibitem{gai10}
Gai, P., Kapadia, S.,
2010.
Contagion in financial networks.
\newblock{Proc. Roy. Soc. A.} {466,} 2401-2423.

\bibitem{}
Gai, P., Haldane, A., Kapadia, S.,
2011.
Complexity, concentration and contagion. 
\newblock{J. Monetary. Econ.} 58(5), 453-470.

\bibitem{}
Gleeson, J.P.,
2013.
Binary-state dynamics on complex networks: pair approximation and beyond. 
\newblock{Phys. Rev. X.} 3, 021004. 

\bibitem{}
Gleeson, J.P., Cahalane, D.J.,
2007.
Seed size strongly affects cascades on random networks.
\newblock{Phys. Rev. E.} 75, 056103.

\bibitem{kobayashi13} 
Kobayashi, T.,
2013.
Network versus portfolio structure in financial systems. 
\newblock{Eur. Phys. J. B.} 86:434.

\bibitem{}
Kobayashi, T., Hasui, K.,
2014.
Efficient immunization strategies to prevent financial contagion.
\newblock{Sci. Rep.} 4, 3834. 

\bibitem{lenzu12}
Lenzu, S., Tedeschi, G.,
2012.
Systemic risk on different interbank network topologies.
\newblock{Physica A.} 391(18), 4331-4341.

\bibitem{lorenz09}
Lorenz, J., Battiston, S., Shweitzer, F.,
2009.
Systemic risk in a unifying framework for cascading processes on networks.
\newblock{Eur. Phys. J. B.} 71(4), 441-460. 

\bibitem{nier07}
Nier, E., Yang, J., Yorulmazer, T., Alentorn, A.,
2007.
Network models and financial stability.
\newblock{J. Econ. Dyn. Cont.} 31(6), 2033-2060.

\bibitem{soramaki07}
Soram\"aki, K., Bech, M.L., Arnold, J., Glass, R.J., Beyeler, W.E.,
2007.
The topology of interbank payment flows.
\newblock{Physica A.} 379(1), 317-333.

\bibitem{upper11} 
Upper, C.,
2011.
Simulation methods to assess the danger of contagion in interbank markets.
\newblock{J. Financ. Stability.} 7(3), 111-125.

\bibitem{watts02}
Watts, D.J.,
2002.
A simple model of global cascades on random networks.
\newblock{Proc. Natl. Acad. Sci. USA.} 99(9), 5766-5771.

\bibitem{}
Watts, D. J., Dodds, P. S.,
2007.
Influentials, networks, and public opinion formation. 
\newblock{J. Consumer. Res.} 34(4), 441-458.

\end{thebibliography}
\end{document}